\nonstopmode

\newcommand{\Cal}[1]{{\cal #1}}

\newcommand{\cbraphi}[1]{\cbra{\Phi#1}}
\newcommand{\cketphi}[1]{\cket{\Phi#1}}

\newcommand{\opHz}{\hat H_0(\eta)}
\newcommand{\opHo}{\hat H_1(\eta)}

\newcommand{\imag}{\textrm{i}}
\newcommand{\diag}{\textbf{diag}}

\newcommand{\eg}{e.g.{}}
\newcommand{\etal}{\textit{et al.}}

\newcommand{\I}[1]{_{\textrm{\scriptsize #1}}}
\newcommand{\U}[1]{\,\textrm{#1}}
\newcommand{\unitmatrix}{\mat{\mathbbm{1}}}
\newcommand{\unitop}{\hat \mathbbm{1}}

\newcommand{\mat}[1]{\hbox{\boldmath{$#1$}\unboldmath}}

\newcommand{\cbra}[1]{\left(\right.\!#1\!\left.\right|}
\newcommand{\cket}[1]{\left|\right.\!#1\!\left.\right)}
\newcommand{\opHeta}{\hat H(\eta)}
\newcommand{\maHeta}{\mat H(\eta)}
\newcommand{\differential}{\>\textrm{d}}
\newcommand{\re}{\textrm{Re\,}}
\newcommand{\im}{\textrm{Im\,}}
\newcommand{\Sum}{\sum\limits}
\newcommand{\Lim}{\lim\limits}

\newcommand{\transpose}{{}^{\textrm{\scriptsize T}}}
\newcommand{\opP}{\hat \Cal P}
\newcommand{\opQ}{\hat \Cal Q}
\newcommand{\maP}{\mat{\Cal P}}
\newcommand{\maQ}{\mat{\Cal Q}}
\newcommand{\opphijket}{|\Phi_j)}
\newcommand{\opphijbra}{(\Phi_j|}
\newcommand{\opPHP}{\opP \opHeta \opP}
\newcommand{\opPHQ}{\opP \opHeta \opQ}
\newcommand{\opQHP}{\opQ \opHeta \opP}
\newcommand{\opQHQ}{\opQ \opHeta \opQ}
\newcommand{\Ejeta}{E(\eta)}
\newcommand{\maQD}{\mat Q(\eta)}
\newcommand{\maQDt}{\mat Q(\eta)\transpose}
\newcommand{\maPHP}{\maP \maHeta \maP}
\newcommand{\maPHQ}{\maP \maHeta \maQ}
\newcommand{\maQHP}{\maQ \maHeta \maP}
\newcommand{\maQHQ}{\maQ \maHeta \maQ}
\newcommand{\oppsijetaket}{|\Psi(\eta))}

\newcommand{\Heff}{\hat H\I{eff}(\eta)}
\newcommand{\maHeff}{\mat{H}\I{eff}(\eta)}
\newcommand{\opphikket}{|\Phi_k)}

\newcommand{\HqqD}{\mat H^{\Cal Q \Cal Q}_0(\eta)}
\newcommand{\HqqN}{\mat H^{\Cal Q \Cal Q}_1(\eta)}
\newcommand{\maE}{\mat E^{(0)}(\eta)}

\documentclass[aps,pra,showpacs,twocolumn]{revtex4-2}
\usepackage{graphicx,subeqnarray,amsfonts,bbm,bm}

\begin{document}
\title{Non-Hermitian Rayleigh-Schr\"odinger Perturbation Theory}
\date{10 December 2003}
\author{Christian Buth}
\email[Corresponding author: ]{Christian.Buth@web.de}
\altaffiliation[Present address: ]{Max-Planck-Institut f\"ur
Physik komplexer Systeme, N\"othnitzer Stra\ss{}e~38,
01187~Dresden, Germany}
\author{Robin Santra}
\altaffiliation[Present address: ]{JILA, University of Colorado,
Boulder, CO 80309-0440, USA}
\author{Lorenz S. Cederbaum}
\affiliation{Theoretische Chemie, Physikalisch-Chemisches
Institut, Ruprecht-Karls-Universit\"at Heidelberg, Im Neuenheimer
Feld~229, 69120~Heidelberg, Germany}

\begin{abstract}
We devise a non-Hermitian Rayleigh-Schr\"odinger perturbation
theory for the single- and the multireference case to tackle both
the many-body problem and the decay problem encountered, for example, 
in the study of electronic resonances in molecules. A complex absorbing
potential~(CAP) is employed to facilitate a treatment of
resonance states that is similar to the well-established bound-state 
techniques. For the perturbative approach, the full CAP-Schr\"odinger
Hamiltonian, in suitable representation, is partitioned according to the
Epstein-Nesbet scheme. The equations we derive in the framework of the 
single-reference perturbation theory turn out to be identical to 
those obtained by a time-dependent treatment in Wigner-Weisskopf 
theory. The multireference perturbation theory is studied for a 
model problem and is shown to be an efficient and accurate method. 
Algorithmic aspects of the integration of the perturbation theories into 
existing \emph{ab initio} programs are discussed, and the simplicity 
of their implementation is elucidated.
\end{abstract}

%
%

%
%
%

\maketitle

\section{Introduction}

Electronic resonances are temporary states which decay by electron emission. 
Prominent examples are Auger decay~\cite{Auger:-23,Burhop:AE-80,
Thompson:AE-85,Tarantelli:RD-92,Buth:IO-03,Buth:IM-03}, electronic
decay of inner-valence ionized clusters~\cite{Cederbaum:GI-97,Zobeley:HE-98,
Santra:ED-99,Santra:ICD-01} or temporary anions~\cite{Jordan:TA-87,Sommerfeld:TA-98,
Wang:PF-00,Sommerfeld:LM-00,Hampe:GS-02,Kalcher:TP-02}.
Electronic resonances are of fundamental importance and their study reveals 
deep physical insights into the complex many-body effects governing molecular
physics (see, for example, Refs.~\cite{Buth:IO-03,Buth:IM-03}).

Resonances cannot be described in terms of bound-state quantum
mechanics because the wave functions are not part of the
$\mathbb L^2$-Hilbert space, since a particle is emitted into the continuum.
Hence, the wave functions are not square-integrable. Associated with each 
resonance is a discrete pole of the $S$ matrix at the complex energy
\begin{equation}
  \label{eq:siegert}
  E\I{res} = E\I{R} - \imag \, \Gamma/2.
\end{equation}
This energy is called Siegert energy~\cite{Gamow:-28,Siegert:DF-39,
Kukulin:TR-89}. $E\I{R}$~is the energetic position of
the resonance state and $\Gamma$~is its decay width.
The Siegert wave function typically displays a localized part, near the decaying
system, resembling a bound state. Asymptotically, however, the Siegert wave function
diverges~\cite{Kukulin:TR-89}.

The discrete nature of a Siegert energy and the quasi-bound behavior of the
corresponding wave function in the vicinity of the decaying system
suggest to employ bound-state techniques in
order to treat resonances quantitatively. This idea is exploited 
by a couple of methods like the stabilization method~\cite{Hazi:SM-70,
Mandelshtam:CD-93}, complex scaling~\cite{Aguilar:-71,Balslev:-71,
Simon:-72,Moiseyev:CS-98}, or the use of a complex absorbing
potential~(CAP)~\cite{Jolicard:-85,Riss:CRE-93,Santra:NH-02}. 
In the latter two approaches, the Siegert energy of a resonance is found
as a discrete eigenvalue of a modified, non-Hermitian Schr\"odinger equation.
The effective Siegert wave function in this eigenvalue problem is
square-integrable, so that bound-state methods can indeed be applied.

Using a CAP, in particular, offers advantages when considering electronic
resonances in molecules or in other extended forms of 
matter~\cite{Santra:NH-02}. Its implementation in \emph{ab initio} packages
for bound-state quantum chemistry is simple and, in some sense, minimally 
invasive. For example, there is no need to modify the calculation of the 
Coulomb integrals. Moreover, in principle any electron-correlation method 
can be combined with a CAP. For example, the CAP method was implemented by 
Sommerfeld~\etal~\cite{Sommerfeld:TA-98} at the multireference 
configuration-interaction level. This program, known as CAP/CI, has already 
found several interesting applications (see, for instance, 
Refs.~\cite{Sommerfeld:LM-00,Santra:EC-01}). More recently, a CAP-based
extension of many-body Green's functions was developed~\cite{Santra:CAP-02}
and used to perform the first \emph{ab initio} determination of 
resonance widths in elastic electron scattering from 
chlorobenzene~\cite{Santra:CAP-03}.

A very promising strategy for applications to large systems is treating the 
CAP-Schr\"odinger eigenvalue problem in electronic configuration space using 
perturbation theory. The non-Hermitian perturbation theory derived in this 
article, simplifies the practical calculation of electronic resonance states 
considerably. We describe the combination of the method of configuration 
interaction with CAPs in Sec.~\ref{sec:MBCAP} and study a 
non-Hermitian single-reference Rayleigh-Schr\"odinger ansatz in 
Sec.~\ref{sec:srpt}. A generalization to a multireference theory 
is presented in Sec.~\ref{sec:mrpt}. In Sec.~\ref{sec:model}, the 
performance of the non-Hermitian multireference perturbation theory 
is demonstrated for a model problem. We summarize in Sec.~\ref{sec:conclusion}.

\section{Complex Absorbing Potentials and Configuration Interaction}
\label{sec:MBCAP}

In order to enable a treatment of decaying states within the framework
of bound-state methods, the molecular system is enclosed by an appropriate
complex potential, which enforces an absorbing boundary condition. This 
artificial potential, called \emph{complex absorbing potential} or CAP, absorbs 
the emitted particle and consequently transforms the former continuum wave 
function into a square-integrable 
one~\cite{Riss:CRE-93,Santra:ED-99,Santra:NH-02}. The CAP is simply added to
the electronic Hamiltonian~$\hat H$ of a molecular system:
\begin{equation}
  \label{eq:CAP-H}
  \opHeta = \hat H - \imag \, \eta \, \hat W \; .
\end{equation}
Here, $\eta$~is a real positive parameter referred to as CAP-strength 
parameter, and $\hat W$~is called CAP operator. It is possible to 
choose~$\hat W$ as a \emph{local one-particle} operator. The exact 
prerequisites that~$\hat W$ must satisfy are derived in 
Ref.~\cite{Riss:CRE-93}. A fairly general, flexible CAP for molecular 
calculations is presented in Refs.~\cite{Santra:ED-99,Santra:EC-01}. 
(See also Sec.~\ref{sec:model}, Eq.~(\ref{eq:CAPmodel}) for a 
typical~$\hat W$.) It should be noted that the Hamiltonian $\opHeta$ 
in~(\ref{eq:CAP-H}) is not Hermitian. Using the usual Hermitian inner 
product, it is not possible to establish an appropriate geometry in the 
eigenspace of $\opHeta$. The natural choice is the complex symmetric bilinear 
form~\cite{Moiseyev:IP-78,Riss:CRE-93,Moiseyev:CS-98,Santra:ED-99,Santra:NH-02}
\begin{equation}
  \label{eq:bilinear-form}
  (\varphi | \psi) := \int\limits_{\mathbb R^3} \varphi(\vec r) \, \psi(\vec r)
  \differential^3 r \; .
\end{equation}

The problem of calculating the complex Siegert energy of a resonance is 
equivalent to solving a complex eigenvalue equation, the
CAP-Schr\"odinger equation~\cite{Riss:CRE-93}
\begin{equation}
  \label{eq:CAP-Schrodinger}
  \opHeta \cket{\Psi(\eta)} = E(\eta) \cket{\Psi(\eta)} \; .
\end{equation}
If Eq.~(\ref{eq:CAP-Schrodinger}) is solved in a complete basis set, 
then there exists an eigenvalue $E(\eta)$ such that the Siegert 
energy is given by~$E\I{res} = \Lim_{\eta \to 0} E(\eta)$. For a finite 
basis set, the limit $\eta \to 0$ merely recovers an approximation to 
the real spectrum of $\hat H$. In this case, an optimal, finite $\eta$ 
is sought, employing the condition~\cite{Riss:CRE-93,Santra:ED-99,Santra:NH-02}
\begin{equation}
  \label{eq:eres}
  \left| \eta \frac{\differential E}{\differential \eta} \right|
  = \hbox{minimum} \; .
\end{equation}

In a practical electron-correlation calculation, a basis 
\begin{equation}
  \label{eq:tlbasis}
    \Cal B = \left\{\cket{\Phi_j}: \ j =1, \ldots, K\right\}
 \end{equation}
of square-integrable many-electron vectors $\cket{\Phi_j}$---often 
eigenvectors of an appropriate Fock operator---is 
introduced~\cite{Szabo:MQC-82}. Typically, the many-electron 
basis is real. The matrix representation of both $\hat H$ and $\hat W$ 
is therefore real; that of $\opHeta$ is complex symmetric.
In other words, Eq.~(\ref{eq:CAP-Schrodinger}) is transformed into a 
complex symmetric matrix eigenvalue problem. See Ref.~\cite{Santra:NH-02} 
for a discussion of some of the spectral properties of complex symmetric 
matrices and of the numerical techniques available for diagonalizing them.
Representing the Hamiltonian of a many-particle system in a systematically 
constructed many-particle basis set and diagonalizing the resulting matrix
is known as the method of \emph{configuration interaction} 
(CI)~\cite{Szabo:MQC-82}. As CI~matrices are in most cases huge, perturbation
theory is often employed to calculate a few selected eigenstates.
In the CAP/CI problem, accelerating the calculation is even more crucial,
since, according to Eq.~(\ref{eq:eres}), finding the Siegert energy of
a resonance requires repeated diagonalization as a function of $\eta$.

\section{Non-Hermitian Single-Reference Perturbation Theory}
\label{sec:srpt}

The Hamiltonian~$\hat H$ of the system is partitioned, following
Epstein and Nesbet~\cite{Chen:PT-02,Epstein:SE-26,Nesbet-55},
into an exact, diagonal part and an off-diagonal perturbation:
\begin{eqnarray}
  \label{eq:epstein}
  \hat H   &=& \hat H_0 + \hat H_1 \nonumber \\
  \hat H_0 &=& \sum_{j} \cketphi{_j}
  \cbraphi{_j} \hat H \cketphi{_j} \cbraphi{_j} \\
  \hat H_1 &=& \sum_{j,k \atop
  j \neq k} \cketphi{_j} \cbraphi{_j}\hat H
  \cketphi{_k} \cbraphi{_k} \nonumber \; .
\end{eqnarray}
The CAP operator~$\hat W$ is partitioned accordingly. Thus,
employing~(\ref{eq:CAP-H}), we obtain
\begin{equation}
  \label{eq:Epstein-Nesbet-CAP}
  \opHeta = \opHz + \opHo \; ,
\end{equation}
which can be used as a starting point for a simple
time-independent Rayleigh-Schr\"odinger perturbation
theory~\cite{Sakurai:MQM-94} based on one
reference~$\cket{\Phi_i}$. Clearly, $\cket{\Phi_i}$ should
approximate the compact bound-state-like part of the resonance
wave function. In a time-dependent framework, $\cket{\Phi_i}$
would describe the discrete, square-integrable state that undergoes decay 
through coupling to the continuum. We refer to $\cket{\Phi_i}$
as the \emph{initial state} to emphasize this fact.

The CAP-Schr\"odinger equation for the unperturbed part
of~(\ref{eq:Epstein-Nesbet-CAP}) is
\begin{equation}
  \opHz \cket{\Phi_j} = E_j^{(0)}(\eta) \cket{\Phi_j} \; ,
\end{equation}
for an arbitrary~$\cket{\Phi_j} \in \Cal B$, where the unperturbed
eigenenergy $E_j^{(0)}(\eta) = \cbraphi{_j} \opHeta \cketphi{_j}$
is the expectation value of the full operator $\opHeta$ with respect
to $\cket{\Phi_j}$. Now the well-known derivation of the 
Rayleigh-Schr\"odinger perturbation series~\cite{Sakurai:MQM-94} can be
transferred to the non-Hermitian case. Up to second order, the
energy of the initial state is given by
\begin{subeqnarray}
  \label{eq:srpt-E}
  \slabel{eq:srpt-E0}
  E_i^{(0)}(\eta) &=& \cbra{\Phi_i} \opHeta \cket{\Phi_i}\\
  \slabel{eq:srpt-E1}
  E_i^{(1)}(\eta) &=& \cbra{\Phi_i} \opHo \cket{\Phi_i} = 0 \\
  \slabel{eq:srpt-E2}
  E_i^{(2)}(\eta) &=& \sum_{f \neq i} \underbrace{\frac{\cbra{\Phi_i} 
  \opHo \cketphi{_f}^2}
  {E_i^{(0)}(\eta) - E_f^{(0)}(\eta)}}_{P_2(f)} \; .
\end{subeqnarray}
Note that in view of Eqs.~(\ref{eq:epstein}) 
and~(\ref{eq:Epstein-Nesbet-CAP}), the first-order 
correction~(\ref{eq:srpt-E1}) vanishes.

To study the real and imaginary parts of the second-order 
correction~(\ref{eq:srpt-E2}), we assume that the matrix elements of
both $\hat H$ and $\hat W$ are real, and we introduce an abbreviated notation: 
\begin{eqnarray*}
  H_{jk} &:=& \cbra{\Phi_j} \hat H \cketphi{_k} \\
  \Delta H_{jk} &:=& H_{jj} - H_{kk}\\
  W_{jk} &:=& \cbra{\Phi_j} \hat W \cketphi{_k} \\
  \Delta W_{jk} &:=& W_{jj} - W_{kk} \; .
\end{eqnarray*}
In the limit $\eta \to 0$, the real part of $P_2(f)$~reduces to
\begin{equation}
  \label{eq:WWRe}
  \lim_{\eta \to 0} \re P_2(f) = \frac{H_{if}^2}{\Delta H_{if}} ,
\end{equation}
which is the well-known Rayleigh-Schr\"odinger result for a
non-degenerate state. The sum in~(\ref{eq:srpt-E2}) becomes an
integral in a complete basis set, and in general the
principal value of this integral must be taken due to the pole
on the real axis.

Now let 
\begin{equation}
  \label{eq:dedef}
\delta_{\varepsilon}(x) = \frac{1}{\pi} \frac{\varepsilon} {x^2 +
  \varepsilon^2} , 
\end{equation}
so that Dirac's delta function can be written as 
\begin{equation}
  \label{eq:Dirac}
  \delta(x) = \lim_{\varepsilon \to 0} \delta_{\varepsilon}(x).
\end{equation}
Equation~(\ref{eq:dedef}) can be used to transform the imaginary part
of $P_2(f)$ into
\begin{equation}
  \label{eq:CAPWig}
  \begin{array}{rcl}
    \im P_2(f) &=& - \pi \, H_{if}^2 \, \delta_{\varepsilon}
		   (\Delta H_{if}) \\
	       &&{}+ 2 \, \pi \, \frac{H_{if} \, W_{if}}{\Delta W_{if}} \,
		   \Delta H_{if} \, \delta_{\varepsilon}(\Delta H_{if}) \\
	       &&{}+ \pi \, \eta^2 \, W_{if}^2 \, \delta_{\varepsilon}
		   (\Delta H_{if})
  \end{array}
\end{equation}
with $\varepsilon := - \eta \, \Delta W_{if}$. The last two terms
in Eq.~(\ref{eq:CAPWig}) represent an artefact introduced by the CAP. 
However, the limit~$\eta \to 0$ implies $\varepsilon \to 0$, so that their 
contribution to $\lim_{\eta \to 0}\im E_i^{(2)}(\eta)$ vanishes. 

As mentioned previously, the initial state $\cket{\Phi_i}$ is spatially
compact. The only purpose of $\hat W$ is to absorb the outgoing wave outside 
the molecular system. It is therefore obvious (see 
Refs.~\cite{Santra:ED-99,Santra:EC-01}) that $\hat W$ can be chosen
in such a way that $W_{if}$ vanishes for any $f$ (including $f=i$). 
Employing such a CAP, we find, even at finite $\eta$, 
\begin{equation}
  \label{eq:WWIm}
  \im P_2(f) = - \pi \, H_{if}^2 \, \delta_{\varepsilon}(\Delta H_{if}) \; .
\end{equation}

Using Eqs.~(\ref{eq:srpt-E}), (\ref{eq:WWRe}), (\ref{eq:Dirac}), and 
(\ref{eq:WWIm}), the Siegert energy, as obtained in second-order 
 perturbation theory in the limit $\eta \to 0$, reads 
\begin{equation}
  \label{eq:WWSiegert}
   E\I{res} = H_{ii} + \Pr\ \sum_{f \neq i}
  \frac{H_{if}^2}{\Delta H_{if}}
  - \imag \, \pi \, \sum_{f \neq i} H_{if}^2 \, \delta(\Delta H_{if}) .
\end{equation}
The decay width, in particular, is given by 
\begin{equation}
\label{eq:WWdecaywidth}
\Gamma_i = 2 \, \pi \, \sum_{f \neq i} H_{if}^2 \, \delta(\Delta H_{if}) .
\end{equation}
Equation~(\ref{eq:WWSiegert}) for $E\I{res}$ is identical to the Siegert
energy derived in the framework of Wigner-Weisskopf 
theory~\cite{Sakurai:MQM-94,Santra:NH-02,Santra:ED-01}, which utilizes 
time-dependent perturbation theory in second order in combination with the
well-known relation~\cite{Dennery:MP-67}
\begin{equation}
  \frac{1}{x + \imag \, \varepsilon} = \Pr \frac{1}{x}
  - \imag \, \pi \, \delta(x) \; .
\end{equation}

In a finite basis set, treating the delta function in 
Eq.~(\ref{eq:WWdecaywidth}) requires care. The usual approach is the direct
evaluation of Eq.~(\ref{eq:WWdecaywidth}) by exploiting Stieltjes-Chebyshev 
moment theory~\cite{Rescigno:EM-79}, a cumbersome and not always numerically stable
scheme. Non-Hermitian Rayleigh-Schr\"odinger perturbation theory offers a 
computationally powerful alternative: The complex energy 
$E_i^{(2)}(\eta)$ (Eq.~(\ref{eq:srpt-E2})) is easy to calculate as a 
function of $\eta$, and optimization of the complex energy in accordance with 
Eq.~(\ref{eq:eres}) immediately yields a numerical approximation to the
second-order level shift in Eq.~(\ref{eq:WWSiegert}) and to the decay 
width~(\ref{eq:WWdecaywidth}).

The energies $E_i^{(0)}(\eta)$, $E_i^{(1)}(\eta)$, and
$E_i^{(2)}(\eta)$ in (\ref{eq:srpt-E}) are the first three terms 
of a perturbation series expansion of the Siegert energy associated with
$\cket{\Phi_i}$. Basic insight into the convergence behavior of such
a series can be gained by analyzing a two-state model~\cite{Finley:CB-96,Olsen:MP-00}.
If $\hat W$ is chosen appropriately, the complex symmetric Hamiltonian
matrix in a basis consisting of the compact vector $\cket{\Phi_i}$ and 
one diffuse vector $\cket{\Phi_f}$ is simply 
\begin{equation}
  \label{eq:Model}
  {\bm H}(\eta;\lambda) = \left[
    \begin{array}{cc}
      H_{ii} & \lambda \, H_{if} \\
      \lambda \, H_{if} & H_{ff} - \imag \, \eta \, W_{ff}
    \end{array}
  \right],
\end{equation}
where $\lambda$ controls the strength of the perturbation.
The eigenvalues of this matrix are
\begin{eqnarray}
  E_{\pm}(\eta;\lambda) & = & 
  \frac{1}{2} (H_{ii}+H_{ff}- \imag \, \eta \, W_{ff}) \\
  &&{} \pm \sqrt{\frac{1}{4} (\Delta H_{if}+ \imag \, \eta \, W_{ff})^2
  + \lambda^2 \, H_{if}^2} \, . \nonumber 
\end{eqnarray}
The function $E_{\pm}(\eta;\lambda)$ has branch-point singularities, 
with respect to $\lambda$, at the radius
\begin{equation}
  \left|\lambda_{\mathrm{BP}}\right| = 
  \left|\frac{\sqrt{\Delta H_{if}^2 + \eta^2 \, W_{ff}^2}}{2 \, H_{if}}
  \right| .
\end{equation}
The expansion of $E_{\pm}(\eta;\lambda)$ in powers of $\lambda$ converges
at the physically relevant point $\lambda = 1$ if 
$\left|\lambda_{\mathrm{BP}}\right| > 1$. In the case $\eta = 0$, the condition
$\left|\Delta H_{if}\right| > 2 \left|H_{if}\right|$ must therefore be 
satisfied. However, if decay from $\cket{\Phi_i}$ into $\cket{\Phi_f}$ 
is efficient, then we expect the energy difference, $\Delta H_{if}$, between 
the two states to be small in comparison to the coupling strength, $H_{if}$. 
A finite CAP can restore convergence under the condition that 
$\left|\eta \, W_{ff}\right| > 2 \left|H_{if}\right|$. The spectrum of 
the two-state matrix ${\bm H}(\eta;\lambda=1)$ is then nondegenerate,
which ensures the existence of two linearly independent 
eigenvectors~\cite{Santra:NH-02}.

If only a single reference is used to describe the initial state, 
it can happen that the optimal $\eta$ (Eq.~(\ref{eq:eres})) cannot
compensate the quasi-degeneracy of $\cket{\Phi_i}$ with a number
of pseudo-continuum states $\cket{\Phi_f}$. Another reason why 
the single-reference perturbation series can be divergent is strong 
coupling of $\cket{\Phi_i}$ to other compact states that are close
in energy. Such an effect implies strong electron correlation and can 
cause a complete breakdown of the independent-particle 
picture~\cite{Cederbaum:CE-86}. Both insufficiencies can be overcome 
by using the multireference approach that is derived in the ensuing 
Sec.~\ref{sec:mrpt}.

\section{Non-Hermitian Multireference Perturbation Theory}
\label{sec:mrpt}

The degenerate time-independent perturbation theory is a special 
multireference technique: A reference subspace of degenerate
states, belonging to a certain unperturbed energy eigenvalue, is taken.  
Then the representation of the full Hamiltonian in this subspace is 
diagonalized to decouple the states and to yield corrections to the 
energy in first and to the wave function in zeroth order.  If the 
degeneracy is removed in the new eigenvector basis, non-degenerate 
perturbation theory can be applied to each eigenvector to obtain higher-order 
corrections~\cite{Sakurai:MQM-94}.

A general multireference approach can be devised analogously. An
arbitrary set of initial states can be taken. Then one can proceed
as described in the previous paragraph.	To carry out this program, we
harness an effective Hamiltonian formalism~\cite{Lowdin:-62,
Lowdin:BT-62,Lowdin:-65,Davidson:PT-78}, which has proven to be a
versatile tool in many cases, see \eg~Refs.~\cite{Albrecht:AA-00,
Albrecht:LAI-02}.

\subsection{Effective Eigenvalue Problem}

The CAP-Schr\"odinger equation~(\ref{eq:CAP-Schrodinger}) shall be
solved for several complex eigenvalues. In practice, a $K$-dimensional basis $\Cal B$
(Eq.~(\ref{eq:tlbasis})) is used to form a complex symmetric matrix
representation of~(\ref{eq:CAP-Schrodinger}). Now $n$~references are 
selected from $\Cal B$, say, $\opphijket,\ j=1, \ldots, n$ ($n \ll K$). 
Projection operators~\cite{Lowdin:BT-62,
Lindgren:AM-86,Santra:EC-01,Santra:NH-02} onto the reference space and 
its complement space, respectively, are defined as 
\begin{equation}
  \label{eq:PQ}
  \opP = \sum_{j=1}^n \opphijket \opphijbra, \quad
  \opQ = \unitop - \opP = \sum_{j=n+1}^K \opphijket \opphijbra,
\end{equation}
obeying
\begin{equation}
  \label{eq:PQprop}
  \begin{array}{c}
    \opP^2 = \opP, \qquad \opQ^2 = \opQ, \qquad \opP\transpose = \opP, \\
    \opQ\transpose = \opQ, \qquad \opP + \opQ = \unitop,
    \quad \opP \, \opQ = 0 \; .
  \end{array}
\end{equation}
Applying~(\ref{eq:PQ}) and (\ref{eq:PQprop}) to~(\ref{eq:CAP-Schrodinger}) yields
\begin{subeqnarray}
  \slabel{eq:opPHP_PHQ}
  \Ejeta \, \opP \oppsijetaket &=& \opPHP \, \oppsijetaket \\
  &&\hspace{5em} {} + \opPHQ \, \oppsijetaket \nonumber \\
  \slabel{eq:opQHP_QHQ}
  \Ejeta \, \opQ \oppsijetaket &=& \opQHP \, \oppsijetaket \\
  &&\hspace{5em} {} + \opQHQ \, \oppsijetaket \; . \nonumber
\end{subeqnarray}
Equation~(\ref{eq:opQHP_QHQ}) is solved for~$\opQ \oppsijetaket$,
which yields with the help of~(\ref{eq:PQprop})
\begin{equation}
  \label{eq:nmQHP_QHQ}
    \opQ \oppsijetaket  = \hat G(\eta) \, \opQHP \oppsijetaket \; .
 \end{equation}
The operator 
\begin{equation}
  \label{eq:Green}
   \hat G(\eta) := [\Ejeta \, \unitop - \opQHQ]^{-1}
 \end{equation}
is the Green's function~\cite{Szabo:MQC-82,Lindgren:AM-86} of the complement space. It 
has a pole where the exact eigenvalue $\Ejeta$ coincides with an eigenvalue 
of~$\opQHQ$. 

Inserting Eq.~(\ref{eq:nmQHP_QHQ}) into~(\ref{eq:opPHP_PHQ}) results in an 
effective eigenvalue problem~\cite{Feshbach:-58,Feshbach:OM-58,Lowdin:-62,
Feshbach:-62,Lindgren:AM-86}
\begin{subeqnarray}
  \label{eq:eff_eigen}
  \slabel{eq:E_eff}
  \Heff \opP \oppsijetaket & = & \Ejeta \opP \oppsijetaket \\
  \slabel{eq:H_eff}
  \Heff &=& \opPHP + \opPHQ \\ 
  &&{} \times \hat G(\eta) \, \opQHP \nonumber
\end{subeqnarray}
for the exact eigenvalue~$\Ejeta$ associated with the 
eigenstate~$\oppsijetaket$. Equation~(\ref{eq:eff_eigen}) is no
simplification but a convenient reformulation of the original
problem, Eq.~(\ref{eq:CAP-Schrodinger}). Note that the matrix 
representation of, \eg, $\opPHP$ is a~$K \times K$ matrix~$\maPHP$ 
with a nonzero~$n \times n$ submatrix.  For notational brevity, the 
$K \times K$ matrix~$\maPHP$ is identified with the smaller nonzero 
$n \times n$ matrix, i.e. with the representation of 
$\opPHP$ in the reference subspace. 

The eigenstate~$\oppsijetaket$ can be obtained from its projection
onto the reference space, $\opP \oppsijetaket$, by using
\begin{equation}
  \oppsijetaket = \opP \oppsijetaket + \opQ \oppsijetaket \; .
\end{equation}
Inserting Eq.~(\ref{eq:nmQHP_QHQ}) yields
\begin{equation}
  \label{eq:wfpt}
    \oppsijetaket = \opP \oppsijetaket + \, \hat G(\eta) \, 
                            \opQHP \, \oppsijetaket \; .
\end{equation}
The second term in this equation represents corrections to $\opP \oppsijetaket$
that arise from coupling to the complement space.

\subsection{Series Expansion}

The matrix representation of the Hamiltonian in the reference
space, $\maPHP$, can be diagonalized to decouple the reference 
configurations. To this end, the complex symmetric eigenvalue problem
\begin{equation}
  \begin{array}{rcl}
    \maPHP \maQD & = & \maQD \maE \\
    \maE & = & \diag(E_1^{(0)}(\eta), \ldots,    E_n^{(0)}(\eta)) 
  \end{array}
\end{equation}
has to be solved. If $\maPHP$ is diagonalizable~\cite{Moiseyev:CS-98,Santra:NH-02},
then $\maQD$ is a complex orthogonal matrix, satisfying~$\maQD \maQDt = 
\maQDt \maQD = \unitmatrix$. In this case, $\maQD$ allows us to perform 
a transformation from $\Cal B = \left\{\opphijket: \ j =1, \ldots K\right\}$ 
to a more useful orthonormal basis:
\begin{equation}
  \label{eq:Qphi}
  |\varphi_j(\eta)) = \cases{\Sum_{k=1}^n Q_{kj}(\eta) \>
  \opphikket	      & ; $j \in \{1  , \ldots, n\}$ \cr
  \qquad \opphijket & ; $j \in \{n+1, \ldots, K\}$ \cr} .
\end{equation}

Let 
\begin{equation}
\check{H}_{jk}(\eta) := (\varphi_j(\eta)| \hat{H}(\eta) | \varphi_k(\eta))
\end{equation}
denote the matrix elements of $\hat{H}(\eta)$ with respect to the new 
basis~(\ref{eq:Qphi}). We observe that 
\begin{equation}
\check{H}_{jk}(\eta) = E_j^{(0)}(\eta) \, \delta_{jk}, \qquad 
j,k = 1, \ldots, n \, ,   
\end{equation}
and apply to the full matrix $\mat{\check H}(\eta)$ Epstein-Nesbet 
partitioning~\cite{Chen:PT-02,Epstein:SE-26,Nesbet-55} into diagonal
and off-diagonal parts:
\begin{equation}
  \label{eq:H_eff_pt}
  \mat{\check H}(\eta) = \mat H_0(\eta) + \lambda \, \mat H_1(\eta) \, .
\end{equation}
The parameter~$\lambda$ is introduced, as in Eq.~(\ref{eq:Model}), to allow
for a systematic perturbation expansion (Sec.~\ref{seq:Approx}). 
Employing Eq.~(\ref{eq:H_eff_pt}), the effective Hamiltonian, Eq.~(\ref{eq:H_eff}), 
represented in the new basis, Eq.~(\ref{eq:Qphi}), reads
\begin{equation}
  \label{eq:H_eff_final}
  \begin{array}{rcl}
  \maHeff &=& \maE + \lambda^2 \, \maQDt \, \maPHQ \\ 
  &&{} \times \mat G(\eta) \, \maQHP \, \maQD \, .
  \end{array}
\end{equation}
With the definition
\begin{equation}
  \Psi_k(\eta) := (\varphi_k(\eta)| \Psi(\eta)), \qquad k=1,\ldots,K \, ,
\end{equation}
the complex eigenvalue problem of~$\maHeff$ can be written as 
\begin{equation}
  \label{eq:E_eff_final}
  \sum_{l=1}^{n} (\maHeff)_{kl} \Psi_l(\eta) = \Ejeta \Psi_k(\eta),
  \quad k=1,\ldots,n  \, .
\end{equation}

In order to solve Eqs.~(\ref{eq:H_eff_final}) and (\ref{eq:E_eff_final}), 
the matrix $\mat G(\eta)$ of the Green's function $\hat G(\eta)$, 
Eq.~(\ref{eq:Green}), must be evaluated. We 
set~$\HqqD := (\maQHQ)\I{diagonal}$ and $\HqqN :=
(\maQHQ)\I{off-diagonal}$, so that 
\begin{equation}
  \label{eq:G_lambda}
  \mat G(\eta) = [E(\eta) \, \unitmatrix - \HqqD - \lambda
  \HqqN]^{-1} .
\end{equation}
The eigenvector $\left(\Psi_1(\eta),\ldots,\Psi_K(\eta)\right)$ 
of $\mat{\check H}(\eta)$ converges, in the limit $\lambda \rightarrow 0$, to 
a Cartesian unit vector $\hat{e}_j \in \mathbb{C}^K$ for some fixed~$j$:
\begin{equation}
  \label{eq:PsiConv}
  \Lim_{\lambda \to 0} \Psi_k(\eta) = \delta_{jk} \, .
\end{equation}
Let us assume the reference space is chosen such that $j \in \{1,\ldots,n\}$.
Then,
\begin{equation}
  \label{eq:EEj}
  \Lim_{\lambda \to 0} E(\eta) = E_j^{(0)}(\eta) \, .
\end{equation}
Provided that
\begin{equation}
  \label{eq:series_conv}
  \left \|\mat g(\eta) \, \left\{(\Ejeta - E_j^{(0)}(\eta)) \, \unitmatrix
  - \lambda \, \HqqN\right\}\right\| < 1 
\end{equation}
holds, where
\begin{equation}
\mat g(\eta) := [E_j^{(0)}(\eta) \, \unitmatrix -
\HqqD]^{-1} \, , 
\end{equation}
$\mat G(\eta)$~in Eq.~(\ref{eq:G_lambda}) can be expanded in a geometric series
(according to Lemma~2.3.3 in Ref.~\cite{Golub:MC-89}, which also holds for 
complex matrices):
\begin{widetext}
  \begin{equation}
    \label{eq:GreensGeom}
    \mat G(\eta) = \Bigl[ \, \Sum_{k=0}^{\infty} (-1)^k \, 
    \left(\mat g(\eta) \, 
    \left\{(E(\eta) - E_j^{(0)}(\eta)) \, \unitmatrix 
    - \lambda \, \HqqN\right\}\right)^k \, \Bigr] \, \mat g(\eta) \; .
  \end{equation}
\end{widetext}
Condition~(\ref{eq:series_conv}) is, of course, satisfied in the limit
$\lambda \rightarrow 0$, in view of Eq.~(\ref{eq:EEj}). At $\lambda = 1$, 
(\ref{eq:series_conv}) implies that the series (\ref{eq:GreensGeom}) 
converges only if the unperturbed energy $E_j^{(0)}(\eta)$ is well
separated from all diagonal elements of~$\maQHQ$. This can, in principle,
always be enforced by choosing a sufficiently large number of references.

\subsection{Approximation}
\label{seq:Approx}

The matrix~$\mat G(\eta)$ in Eq.~(\ref{eq:GreensGeom}) still depends on
the exact energy~$E(\eta)$. Equations~(\ref{eq:H_eff_final}), 
(\ref{eq:E_eff_final}), and (\ref{eq:EEj}) allow us to make the following
perturbative ansatz:
\begin{equation}
  E(\eta) = E^{(0)}_j(\eta) + \Sum_{i=2}^4 \lambda^i \, E^{(i)}(\eta) + 
  O(\lambda^5) \, .
\end{equation}
Thus, an expansion of the Green's function in terms of $\lambda$,
\begin{equation}
  \label{eq:Grexp}
  \mat G(\eta) = \Sum_{i=0}^2 \lambda^i \, \mat G^{(i)}(\eta) + O(\lambda^3) \, ,
\end{equation}
can be determined:
\begin{eqnarray}
  \mat G^{(0)}(\eta) &=& \mat g(\eta) \nonumber \\
  \mat G^{(1)}(\eta) &=& \mat g(\eta) \, \HqqN \, \mat g(\eta) 
    \nonumber \\
  \label{eq:Gpt}
  \mat G^{(2)}(\eta) &=& \mat g(\eta) \, \HqqN \, \mat g(\eta) \, 
    \HqqN \, \mat g(\eta) \\
  &&{}-E^{(2)}(\eta) \, \mat g(\eta)^2 \, . \nonumber
\end{eqnarray}
Utilizing Eqs.~(\ref{eq:H_eff_final}) and (\ref{eq:Gpt}), a perturbation
series for the effective Hamiltonian is obtained, which is inserted into 
the effective eigenvalue problem~(\ref{eq:E_eff_final}), together with the 
expansion (see Eq.~(\ref{eq:PsiConv}))
\begin{equation}
  \label{eq:Psiexp}
  \Psi_k(\eta) = \delta_{jk} + \Sum_{i=1}^2 \lambda^i \Psi^{(i)}_k(\eta)
  + O(\lambda^3) \, .
\end{equation}
We assume intermediate normalization, so that 
\begin{equation}
  \label{eq:Psiint}
  \Psi^{(i)}_j(\eta) = 0 
\end{equation}
for all $i\ge 1$. Upon sorting for orders in $\lambda$, the energy corrections, 
up to fourth order, are found to be
\begin{widetext}
  \begin{subeqnarray}
    \label{eq:MRenergy}
    E^{(2)}(\eta) & = & \sum_{k=n+1}^K \frac{\check
      H_{jk}(\eta)^2}{E_j^{(0)}(\eta) - \check H_{kk}(\eta)} \slabel{eq:MR2} \\
    E^{(3)}(\eta) & = & \sum_{k,l=n+1 \atop k \neq l}^K 
      \frac{\check H_{jk}(\eta) \,
      \check H_{kl}(\eta) \, \check H_{lj}(\eta)}{(E_j^{(0)}(\eta) - 
      \check H_{kk}(\eta)) \, (E_j^{(0)}(\eta) - \check H_{ll}(\eta))} \\
    E^{(4)}(\eta) & = & \sum_{k=1 \atop k \neq j}^n \sum_{l=n+1}^K
      \frac{\check H_{kl}(\eta)^2 \, \check H_{lj}(\eta)^2}{(E_j^{(0)}(\eta) -
      E_k^{(0)}(\eta)) \, (E_j^{(0)}(\eta) - \check H_{ll}(\eta))^2} \\
    &&{}- E^{(2)}(\eta) \, \sum_{k=n+1}^K \frac{\check H_{jk}(\eta)^2}
      {(E_j^{(0)}(\eta) - \check H_{kk}(\eta))^2} \nonumber \\
    &&{} + \sum_{k,l=n+1 \atop k \neq l}^K \frac
      {\check H_{jk}(\eta)^2 \, \check H_{kl}(\eta)^2}
      {(E_j^{(0)}(\eta) - \check H_{kk}(\eta))^2
      \, (E_j^{(0)}(\eta) - \check H_{ll}(\eta))} \, . \nonumber
  \end{subeqnarray}
\end{widetext}

Chen~\etal{} have independently derived a formally similar theory in 
Ref.~\cite{Chen:PT-02}, using a direct Taylor series expansion 
of the secular equation for the energy of bound states
of a molecular system. They provide explicit expressions
for the perturbation expansion of the real energy up to third order. Chen~\etal{}
further observe that this kind of multireference approach reduces
to single-reference Epstein-Nesbet perturbation theory if only one
reference is used. Of course, the non-Hermitian multireference perturbation
theory of this section also reduces to the non-Hermitian single-reference 
theory.

From the perturbative treatment of Eq.~(\ref{eq:E_eff_final}) we further 
conclude, for $k=1,\ldots,n$ ($k\neq j$), that
\begin{widetext}
  \begin{subeqnarray}
    \label{eq:psiref}
    \Psi_k^{(1)}(\eta) &=& 0 \\
    \Psi_k^{(2)}(\eta) &=& \sum_{l=n+1}^K \frac{\check H_{kl}(\eta) \, 
        \check H_{lj}(\eta)}{(E^{(0)}_j(\eta) - E^{(0)}_k(\eta)) \, 
        (E_j^{(0)}(\eta) - \check H_{ll}(\eta))} \, .
  \end{subeqnarray}
\end{widetext}
These results represent corrections to the wave function in the
reference space. Corrections to the eigenvector that affect only 
components with respect to the complement space can be obtained 
from~(\ref{eq:wfpt}). Hence, the components $\Psi_k(\eta)$, $k=n+1,\ldots,K$,
are given by 
\begin{equation}
  \label{eq:wfG}
  \Psi_k(\eta) = \lambda \, \sum_{l=1}^n \left(\mat G(\eta) \,
  \maQHP \, \maQD\right)_{kl} \, \Psi_l(\eta) \, .
\end{equation}
Using Eqs.~(\ref{eq:Grexp}), (\ref{eq:Gpt}), (\ref{eq:Psiexp}),
(\ref{eq:Psiint}), and (\ref{eq:psiref}), the first- and second-order 
corrections of the eigenvector in the complement space are as follows 
($k=n+1,\ldots,K$):
\begin{widetext}
  \begin{subeqnarray}
    \label{eq:MRWF}
    \Psi_k^{(1)}(\eta) &=& \frac{\check H_{kj}(\eta)}
      {E_j^{(0)}(\eta) - \check H_{kk}(\eta)} \\
    \Psi_k^{(2)}(\eta) &=&
      \sum_{l=n+1 \atop l \neq k}^K
      \frac{\check H_{kl}(\eta) \, \check H_{lj}(\eta)}
      {(E_j^{(0)}(\eta) - \check H_{kk}(\eta)) \, 
      (E_j^{(0)}(\eta) - \check H_{ll}(\eta))} \, .
   \end{subeqnarray}
\end{widetext}
These corrections to the wave function can also be useful for 
the real case ($\eta=0$) since they are not explicitly given 
in Ref.~\cite{Chen:PT-02}.

\subsection{Computational Algorithm}

In iterative eigenvalue solvers, as for example the
Lanczos~\cite{Lanczos:IM-50,Cullum:LA-85,Golub:MC-89} or the 
Davidson~\cite{Davidson:IC-75} algorithm, a matrix-vector multiplication
is carried out in each iteration. This is the slowest step in these 
algorithms, and it therefore determines their overall performance. The 
following sketch of an algorithm for the fourth-order treatment of the
Siegert energy, Eq.~(\ref{eq:MRenergy}), reveals that this perturbative 
approach is roughly as expensive as a \emph{single} iteration of an 
iterative block-eigenvalue solver.

\begin{enumerate}
  \item Choose a set of $n$ references. Perform a full diagonalization
        of the complex symmetric matrix representation of $\hat{H}(\eta)$ 
        in the reference space ($H_{kl}(\eta)$, $k,l = 1,\ldots,n$) to 
        obtain $\mat E^{(0)}(\eta)$ and $\mat Q(\eta)$. A highly efficient
	program for this purpose is described in Ref.~\cite{Bar-On:FD-97}.
  \item Determine all many-particle configurations, beyond the reference
        space, that contribute to the energy in (\ref{eq:MRenergy}). Since, 
	for electronic-structure problems, $\hat{H}(\eta)$ consists of
	one- and two-body operators, a consistent treatment in fourth order
	requires the inclusion of single, double, triple, and quadruple
	excitations of the references. These excitations form the complement
	space and define, together with the references, the total vector 
	space dimension, $K$. Calculate and store all diagonal matrix 
	elements of $\hat{H}(\eta)$ with respect to the complement space:
	$d_k(\eta) := H_{kk}(\eta)$,  $k = n+1,\ldots,K$. 
  \item Let $E_j^{(2)}(\eta) = E_j^{(4,1)}(\eta) = E_j^{(4,2)}(\eta) = 0$,
        $j=1,\ldots,n$. Loop over~$k$ ($k=n+1, \ldots,K$):
    \begin{enumerate}
      \item Transform column~$k$ of the coupling block:
	\[\check H_{jk}(\eta) = \Sum_{l = 1}^n Q_{lj}(\eta) 
	H_{lk}(\eta),  \ j = 1, \ldots, n\]
      \item For $j=1,\ldots,n$ do
        \[E_j^{(2)}(\eta) = E_j^{(2)}(\eta) + \frac{\check H_{jk}(\eta)^2}
            {E_j^{(0)}(\eta) - d_k(\eta)}\]
	$$\begin{array}{rcl}
          E_j^{(4,1)}(\eta) &=& E_j^{(4,1)}(\eta) + \frac{\check H_{jk}(\eta)^2}
	  {(E_j^{(0)}(\eta) - d_k(\eta))^2} \\
          &&{} \times \Sum_{l=1 \atop l \neq j}^n
	  \frac{\check H_{lk}(\eta)^2}{E_j^{(0)}(\eta) - E_l^{(0)}(\eta)}
        \end{array}$$  
	\[E_j^{(4,2)}(\eta) = E_j^{(4,2)}(\eta) + \frac{\check H_{jk}(\eta)^2}
            {(E_j^{(0)}(\eta) - d_k(\eta))^2}\]    
      \item Save the elements $E_j^{(2)}(\eta)$, $E_j^{(4,1)}(\eta)$,
            $E_j^{(4,2)}(\eta)$, and, if possible, $\check H_{jk}(\eta)$.
    \end{enumerate}
  \item Let $E_j^{(3)}(\eta) = E_j^{(4,3)}(\eta) = 0$, $j=1,\ldots,n$.
        Loop over $k$ from $n+1$ to $K$, for each $j \in \{1,\ldots,n\}$:
    \begin{enumerate}
      \item \[S_k(\eta) = \Sum_{l=n+1}^K 
            \frac{\check H_{jl}(\eta) H_{lk}(\eta) \,
	    (1 - \delta_{kl})}{E_j^{(0)}(\eta) - d_l(\eta)}\]
	    \[T_k(\eta) = \Sum_{l=n+1}^K
            \left(\frac{\check H_{jl}(\eta) H_{lk}(\eta) \,
            (1 - \delta_{kl})}{E_j^{(0)}(\eta) - d_l(\eta)}\right)^2\]
      \item \[E_j^{(3)}(\eta) = E_j^{(3)}(\eta) +
              \frac{\check H_{jk}(\eta) S_k(\eta)}
              {E_j^{(0)}(\eta) - d_k(\eta)}\]
	    \[E_j^{(4,3)}(\eta) = E_j^{(4,3)}(\eta) +
              \frac{T_k(\eta)}{E_j^{(0)}(\eta) - d_k(\eta)}\]  
    \end{enumerate}
  \item 
      $$\begin{array}{rcl}
        E_j(\eta) &=& E_j^{(0)}(\eta) + E_j^{(2)}(\eta) + E_j^{(3)}(\eta) 
          + E_j^{(4,1)}(\eta) \\
          &&{} - E_j^{(2)}(\eta) E_j^{(4,2)}(\eta) + E_j^{(4,3)}(\eta)
      \end{array}$$
\end{enumerate}

The algorithm requires that at least two vectors of the dimension of the 
complement space can be held in main memory. If more memory
is available (core memory or hard disk), at least some of the transformed
matrix elements $\check H_{jk}(\eta)$ should also be kept. Otherwise, all
of them have to be recalculated in step 4. The matrix $H_{kl}(\eta)$, 
$k,l = n+1,\ldots,K$, in the complement space is in general much too big to be
stored (except for the diagonal elements $d_k(\eta)$), so that recalculation 
is the only practical option.

For a second-order treatment, the algorithm simplifies considerably. 
Only steps 1, 2, 3(a), and the step affecting $E_j^{(2)}(\eta)$ in 3(b)
are needed. The memory requirements are minimal. The second-order perturbation
theory can thus be applied to large systems not amenable to a solution by an 
iterative diagonalization procedure.

Note that in second order only single and double excitations of the
reference configurations play a role. This is also true in third order. In third 
order, however, in addition 3(c) and the first halves of 4(a) and 4(b) have to be 
carried out. Step 4(a) implies the multiplication of the Hamiltonian matrix
in the complement space with a vector (for each $j$). This increases the 
computational effort significantly and is comparable to a single step in an 
iterative block diagonalization procedure. As mentioned previously, in the
fourth-order formalism, triple and quadruple excitations add a new level of 
complexity.

Recall that the CAP strength $\eta$ must be optimized in order to fulfill
Eq.~(\ref{eq:eres}) for those resonances that can be approximated well within 
the reference space. The second-order approach is exceptionally simple.
It should therefore be employed to identify potential resonances and to 
preoptimize $\eta$ for each of them. Then, the expensive step 4 can be 
restricted to a few selected $j \in \{1,\ldots,n\}$ and to a relatively
small $\eta$ range. In this way, the inclusion of higher-order corrections 
to the Siegert energies can be achieved particularly efficiently.

\section{Model Problem}
\label{sec:model}

A model problem is studied in this section, which serves to illustrate the 
multireference perturbation theory derived in Sec.~\ref{sec:mrpt}. We consider 
$s$-wave scattering of an electron from the spherically symmetric 
potential~\cite{Santra:NH-02}
\begin{equation}
  \label{eq:Vmodel}
   V(r) = \cases{	  -V_0 & ; $0 \leq r < a $ \cr
		\phantom{-}V_0 & ; $a \leq r < 2a$ \cr
		\phantom{-} 0  & ; $r \geq 2a	 $ \cr} .
\end{equation}
The effective one-dimensional Hamiltonian for this problem reads, in atomic units,
\begin{equation}
  \label{eq:Hmodel}
  \hat H = -\frac{1}{2}\frac{\differential^2}{\differential r^2} + V(r) \, .
\end{equation}
There exists a quasi-analytic solution~\cite{Santra:NH-02}, giving $-6.353803650$ for 
the only bound state and
\begin{equation}
  \label{eq:exactenergy}
  E\I{1st} = 4.001414397 - 0.003616371 \, \imag 
\end{equation}
for the first resonance, assuming~$V_0 = 10,\ a = 1$.

In the following we employ the CAP operator
\begin{equation}
  \label{eq:CAPmodel}
  \hat W = \cases {(r - c)^2 & ; $r \geq c$	    \cr
		   \quad   0 & ; $0 \leq r < c$ \cr} .
\end{equation}
The parameter~$c$ defines a sphere about the origin, inside of which the CAP does 
not affect the wave function.  If~$c$ is too large, with respect to a chosen finite 
basis set, then an outgoing wave will be reflected by the basis set wall and will not
be absorbed by the CAP. Conversely, if~$c$ is too small, the perturbation of the wave 
function near the origin is too large. A good choice is 
$c = 2a = 2$~\cite{Santra:NH-02}. 

The CAP Hamiltonian $\hat{H}(\eta)$ of the 
model problem is given by Eqs.~(\ref{eq:CAP-H}), (\ref{eq:Hmodel}), and (\ref{eq:CAPmodel}).
The matrix representation of $\hat{H}(\eta)$ is constructed using a basis of 
particle-in-a-box functions~($k = 1, \ldots, K$)
\begin{equation}
  \label{eq:model_basis}
  \phi_k(r) = \cases{
  \sqrt{\frac{2}{L}} \sin \left(\frac{k\pi r}{L} \right) & ; $0 \leq r < L$ \cr
  \qquad 0						 & ; $r \geq L$	    \cr} \; .
\end{equation}
A spatial extension of the basis set of~$L = 10$ is assumed throughout, which 
is sufficiently large compared with~$c = 2$. Furthermore, a basis dimension 
of~$K=5000$ is chosen.

A particle-in-a-box function is not a good reference, because its overall shape differs 
substantially from that of a resonance wave function~\cite{Santra:NH-02}.
Hence, the overlap between the best references and the resonance itself is small.
The matrix representation in the basis~(\ref{eq:model_basis}) therefore has little
similarity with a CI matrix. An improved matrix representation of the real 
Hamiltonian~(\ref{eq:Hmodel}) can be created by diagonalizing the matrix representation
of~(\ref{eq:Hmodel}) in a basis of particle-in-a-box functions~(\ref{eq:model_basis}) 
for a potential well depth of~$V_0 = 20$. The $n$ energetically lowest 
eigenvectors in this modified potential serve as references. The matrices of the real and 
imaginary parts for~$V_0 = 10$ are then formed with the help 
of~(\ref{eq:model_basis}) and projected onto the real eigenbasis calculated 
with~$V_0 = 20$. This procedure yields matrices that are nearly diagonally 
dominant, and thus they bear more resemblance to CI matrices. Especially 
the near diagonal dominance of the matrix representation assures a sufficient
overlap of the best references with the first resonance of the model problem.
Nevertheless, \emph{single-reference} perturbation theory is not adequate for this example.

\begin{figure}
  \begin{center}
    \includegraphics[width=8.5cm,clip]{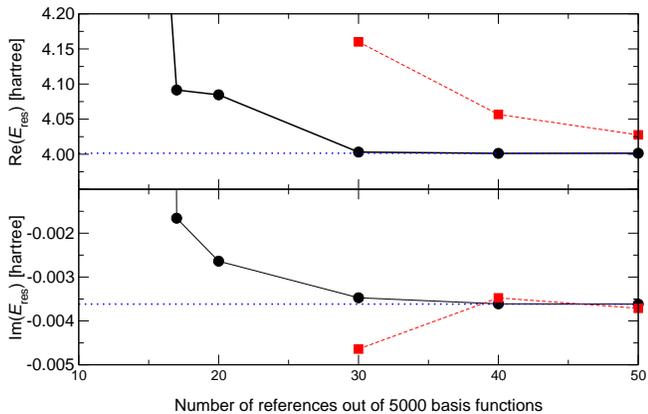} 
    \caption{(Color online) The complex Siegert energy of the energetically
             lowest resonance in the model potential, Eq.~(\ref{eq:Vmodel}),
             for an increasing number of references. The resonance energy 
             is computed (a) by solving the CAP-Schr\"odinger eigenvalue
	     problem just within the reference space (squares in the figure) 
	     and (b) by non-Hermitian multireference perturbation theory up to
             second order (circles). The real part of the Siegert energy
	     is shown in the upper panel, the imaginary part in the lower one.
	     The dotted horizontal lines indicate the exact 
	     values~(\ref{eq:exactenergy}).} 
    \label{fig:conv}
  \end{center}
\end{figure}

In order to demonstrate the usefulness of multireference perturbation theory 
in the non-Hermitian case, we compute the Siegert energy of the first resonance
in two ways. First, we diagonalize $\hat{H}(\eta)$ in the reference space
and employ Eq.~(\ref{eq:eres}) without any perturbative correction. It turns
out that in this case more than 20 references are needed to satisfy the optimization
criterion~(\ref{eq:eres}). Second, we apply the second-order scheme, 
Eq.~(\ref{eq:MR2}). The results are compared in Fig.~\ref{fig:conv}, as a function 
of the number of references, $n$. The real and imaginary parts of the computed Siegert 
energy are shown separately. The dotted horizontal lines in the figure indicate the exact 
values~(\ref{eq:exactenergy}). It is evident that the convergence of the Siegert energy with 
increasing $n$ can be accelerated considerably by the application of perturbation theory. 
Not only is it possible, in the second-order approach, to satisfy Eq.~(\ref{eq:eres}) with 
a relatively small number of references, the accuracy at a given $n$ is also significantly 
improved.

In this example, the application of multireference perturbation theory
takes about~$80 \U{s}$ for 40~references and 70~steps with varying~$\eta$. 
Carrying out a full diagonalization for 70~different $\eta$ values takes 
approximately~$20 \U{d}$ on the same computer.

\section{Conclusion}
\label{sec:conclusion}

Non-Hermitian Rayleigh-Schr\"odinger perturbation theories, based on
Epstein-Nesbet partitioning, have been devised and examined in this
article. The single- and the multireference theory are interesting tools
for investigating the decay properties of resonance states---which represent
the most conspicuous manifestation of quantum effects in scattering processes.
Use of a complex absorbing potential is made in order to transform the time-dependent
decay problem into a non-Hermitian, time-independent, bound-state-like problem that 
can be treated with~$\mathbb L^2$~techniques.

In a complete basis set, the single-reference perturbation theory
turns out to lead to equations identical to those obtained in the framework of 
Wigner-Weisskopf theory. However, the non-Hermitian theory offers computational 
advantages that should be exploited.

Flexibility can be added by using more than one reference. We have developed the
non-Hermitian multireference perturbation theory through fourth order with respect to 
the energy and through second order for the wave function. The application to a model
problem illustrates the efficiency and accuracy of the method.

The integration of non-Hermitian perturbation theory into existing \emph{ab initio} 
programs for solving the electronic-structure problem in finite systems is straightforward 
and is presently carried out utilizing the multireference configuration-interaction program 
\textsc{diesel}~\cite{Hanrath:NA-97,Hanrath:IS-99,Hanrath:CI-00}. The matrix elements 
of the CAP operator can be obtained efficiently~\cite{Santra:EC-01}, with only minor 
modification of the \textsc{diesel}~program. The second-order non-Hermitian multireference 
perturbation theory, in particular, may open the door to a theoretical understanding of
electronic resonance physics in large molecular systems.

\begin{acknowledgments}
We would like to thank Imke B. M\"uller and Thomas Sommerfeld for
helpful discussions. R.~S.~and L.~S.~C.~gratefully acknowledge
financial support by the Deutsche Forschungsgemeinschaft~(DFG).
\end{acknowledgments}

\bibliography{Literatur,solid,cbuth}
\end{document}